\documentclass[a4paper]{amsart}             
\usepackage{amssymb}               
\usepackage{verbatim}             
\usepackage{graphicx}    
\usepackage{upref}    
\usepackage[latin1]{inputenc}  

\makeatletter             
\let\latex@makecaption\@makecaption             
\def\@makecaption{\small\latex@makecaption}             
\makeatother              
\allowdisplaybreaks

\begin{document}                
 \title[ Some properties of Euler capital allocation ]{Some properties of Euler capital allocation  }                 

\author[L.\ Holden]{Lars Holden}                   
\address{Norwegian Computing    
Center, P.\ O.\ Box 114 Blindern,                   
NO--0314 Oslo, Norway}                   
\email{lars.holden@nr.no}                   
 \date{\today\ } 
\maketitle     

%
%

\begin{abstract}   
The paper discusses capital allocation using the Euler formula and
focuses on the risk measures Value-at-Risk (VaR) and Expected
shortfall (ES). Some new results connected to this capital allocation
is shown. Two examples illustrate that capital allocation with VaR is
not monotonous which may be surprising since VaR is monotonous. 
A third example illustrates why the same risk measure should be used in
capital allocation as in the evaluation of the total portfolio.
 We show how simulation may be used in order to estimate the expected
Return on risk adjusted capital in the commitment period of an asset.
Finally, we show how Markov chain Monte Carlo may be used in the
estimation of the capital allocation.
\end{abstract}                   


                   
                  
                  
\section{Introduction}       \label{intro}                
   
The regulatory framework requires a quantification of the total risk  
in a corporate. Based on the quantification and a risk measure there  
is a requirement for an economic capital that is able to absorb  
potential losses. The competition makes it necessary for a financial  
corporate to ensure an efficient use of their economic capital. As a  
part of improving the use of the economic capital, the capital that is  
allocated to each asset is calculated. This makes it possible to  
evaluate each part of the portfolio. This paper studies some  
properties of capital allocation assuming that the stochastic  
properties of the different assets including their correlations are  
known. Hence, we will not discuss how to estimate these distributions.  
  
The two papers Artzner et al. \cite{Artzner97} and \cite{Artzner99}  
have initiated a large number of papers giving a much better  
understanding of capital allocation. There seem to be an agreement  
that capital should be allocated proportional with the partial  
derivatives of the risk measure and use the Euler Theorem, see Tasche  
\cite{Tasche04}. This was first proposed by Litterman \cite{Litterman96}.  
 Denault \cite{Denault01} argues for the same capital  
allocation using theory from cooperative games. Kalkbrener et al.,  
\cite{Kalkbrener04} shows how axioms assuming that the  
capital allocation is linear and diversifying lead to  the same capital  
allocation. Tasche \cite{Tasche06} and \cite{Tasche07} and 
Fischer \cite{Fischer03} give an overview of the argument  
for using Euler allocation.  
  
A large number of authors, see e.g. Acerbie et al \cite{Acerbi02}, argue that the risk  
measure should be coherent. This includes for example expected  
shortfall (ES) but excludes Value-at-Risk (VaR), the most used risk  
measure. VaR is used as a part of the Basel II framework. VaR is not  
subadditive implying that there may be an additional cost when adding  
two portfolios instead of a saving due to diversification. Several  
authors argue that one should use ES which is the smallest measure  
that is law-permitting dominating VaR, see Tasche \cite{Tasche04}.  
  
We will contribute in this area with several different examples. First  
illustrating that capital allocation using VaR is not monotonous.   
This may be surprising since VaR is monotonous. It may be  
a larger problem than the missing subadditivity of VaR in a regular  
use of VaR in capital allocation in corporate. This easily leads to  
suboptimal performance of the management of the portfolio.  
  
The regulatory framework may require that VaR is used for estimating  
the economic capital needed for a portfolio. It may however be  
tempting to use ES for the capital allocation of each asset since this  
is a coherent measure with better mathematical properties. We give an  
example illustrating that the same risk measure should be used for the  
total portfolio and the capital allocation, else one easily gets  
conflict of interest.  
  
Each part of the portfolio may be a commitment for a longer period,  
and the corporate will not always be able to end the part of the  
portfolio that is not cost efficient based on the latest evaluations.  
Hence, a new investment should be evaluated based on the expected  
capital requirement in the commitment period for the investment, not  
only based on today's portfolio. Capital  
allocation based on the present portfolio is not optimal since the  
total portfolio may change during the period where it is not possible  
to change the condition for a part of the portfolio. The natural alternative is  
to simulate the portfolio and use the expected average value of the economic  
capital for the commitment period of the new investment. This is illustrated in an example.  
  
The numerical calculations involved in a capital allocation may be  
very computer intensive. Direct Monte Carlo simulation may not work  
since we need a large number realizations to make a good evaluation of  
the tail behavior. Importance sampling reduces the problem, see  
Kalkbrener et al. \cite{Kalkbrener04}, but it may still be too CPU  
requiring. We propose to use Markov chain Monte Carlo. Then we may  
focus such that all realizations are used in the  
quantification. This is illustrated in an example.  
  
\section{Model}  
In the following we use the terminology from Tasche  
\cite{Tasche07}. Let the random variable $X_i$ denote the cash flow  
from asset $i$. A portfolio $X$ consists of $n$ different assets $$X =  
\sum_{i=1}^n X_i. $$ Further, let $EC(X)= \rho(X)$ denote the economic  
capital that is deemed necessary by the regulator or corporate in  
order to handle a possible unfortunate development of the portfolio.  
$\rho (X)$ is a risk measure on the portfolio $X \in V$ where $V$ is  
the set of real valued random variables. We may define $\rho$ as VaR,  
ES, the standard deviation multiplied with a constant or any other  
measure for the uncertainty. We need to define some properties on risk  
measures.  A risk measure is {\it monotonous}  
if $X,Y  
\in V$ and $X\leq Y$ almost everywhere implies  
 $$\rho(X)\geq \rho(Y),$$   
{\it subadditive } if $X,Y  
\in V$ implies $$\rho(X+Y)\leq \rho(X) + \rho(Y),$$  
 {\it positive  
homogeneous} if $X \in V$, $h \in R$, and $h> 0$ implies   
$$\rho( h X) = h \rho (X)$$  
 and {\it translation invariant} if $X \in V$ and $h \in R$  
implies   
$$\rho(X+h)= \rho(X) - h.$$   
A risk measure is denoted {\it coherent }  
if it is monotonous, subadditive, positive homogeneous and  
translation invariant. In this paper we will assume the risk measure  
is positive homogeneous but do not require the other properties above.  
  
In order to define the derivative of $\rho$ we introduce the function  
$$ f({\bf u}) = \rho( \sum_{i=1}^n u_i X_i) $$ where ${\bf u}= (u_1,  
\cdots , u_n)$.  Assuming the risk measure is positive homogeneous, it  
satisfies the Euler formula, see Tasche \cite{Tasche04}   
$$ f({\bf u }) =  
\sum_{i=1}^d u_i \frac{\partial f}{ \partial u_i} ({\bf u })$$   
where  
we assume the partial derivatives exists. This makes it natural to  
allocate the capital $$ \rho(X_i,X) = \frac{\partial f}{ \partial u_i}  
({\bf u }_1)$$ where ${\bf u }_1= (1,\cdots, 1)$ to the asset $  
X_i$. This capital allocation is denoted Euler allocation. We see  
immediately that this capital allocation is linear,  
e.g. $$\rho(X_i+X_j,X) = \rho(X_i,X) + \rho(X_j,X) $$ and has the full  
allocation property, e.g. $$ \rho(X) = \sum_{i=1}^n \rho(X_i,X). $$  
  
 Assuming $\rho(X)$ is coherent, Denault \cite{Denault01}  
has proved using game theory, that Euler allocation is the only  
allocation satisfying that we always have $\rho(X_i,X)\leq  
\rho(X_i)$. Kalkbrener et al., \cite{Kalkbrener04}, has proved a similar  
result using axioms. This is a critical property closely connected to subadditivity.  
It states that  
an asset is allocated less capital as part of a portfolio than alone.   
Tasche \cite{Tasche03} and \cite{Tasche07} give an overview of the argument for  
this capital allocation.  
  
Another measure on the portfolio is the Return on risk adjusted  
capital, $$ \mbox{RORAC}(X)= \frac{EX}{EC(X)}. $$ For each asset $X_i$  
in a portfolio $X$ we define $\mbox{RORAC}(X_i,X)=EX_i/\rho(X_i,X).$ A  
capital allocation is denoted RORAC compatible, see Tasche \cite{Tasche03}, if there exists  
$\varepsilon_i>0$ such that $ \mbox{RORAC}(X_i,X)> \mbox{RORAC}(X) $  
implies $$ \mbox{RORAC}(X+hX_i)> \mbox{RORAC}(X) $$ for all  
$0<h<\varepsilon_i$. The intuition is that if asset $X_i$ has higher  
RORAC than the entire portfolio, then increasing this asset in the  
portfolio increases the RORAC of the portfolio. This is illustrated in Example 3.

 Tasche \cite{Tasche99} proves that  
Euler allocation is the only allocation that is RORAC  
compatible. In the following two sections we will discuss  
Euler allocation for two of the most popular risk measures.  
  
\section{Value-at-Risk, VaR}  
  
Defined the $\alpha $-quantile $q_{\alpha}$ as  
$$  q_{\alpha}(X) =   
\inf\{ z \in R| P(X\leq z) \geq \alpha  \}.  $$  
Then the  risk measure VaR is defined as $ VaR_{\alpha}(X) =  q_{\alpha}(-X) $.  
We typically have $\alpha \geq 0.99$ implying that estimation of VaR   
only focus on one point in the tail. This makes estimates for VaR much less stable   
than for example standard deviation.   
  
VaR is monotone, positive homogeneous, and translation invariant.   
But it is also well-known that VaR is not subadditive since we may have   
$$ VaR_{\alpha}(X + Y) > VaR_{\alpha}(X) + VaR_{\alpha}(Y). $$  
This implies that VaR is not a coherent risk measure. VaR expresses  
the economical capital necessary to ensure that the probability for a  
default is less than $\alpha$. Hence, VaR focuses only on one  
point in the cumulative distribution of $ X$, the maximum value $z$ where  
 $P(X \geq z) \geq \alpha $. This property is not  
additive. This implies that if we have two portfolios $X $ and  
$Y$ the capital requirement may be larger when we add them  
together than if we keep them separate. This may be an argument for   
splitting the portfolio and hence the corporate  
in two. This may seem counter-intuitive as a risk measure and a   
not wanted property for the corporate for strategic reasons.  
There are several papers stating that risk measures that are not  
coherent should not be used. Artzner et al. \cite{Artzner99} give three examples  
with discrete random variables, one illustrating that the measure is  
not subadditive, another illustrating that VaR fails to recognize  
concentration of risk and fails to encourage a reasonable allocation  
of risk between agents.  
Tasche \cite{Tasche02} gives an example with two independent Pareto   
distributions that does not satisfy subadditivity.  
Acerbi et al, \cite{Acerbi02}, writes `` ..if a measure is not coherent we just choose   
not to call it a risk measure at all'',  
 particular due to the missing subadditive property and since there   
exists coherent risk measures with satisfactory properties.   
  
The Euler allocation for VaR is   
\begin{equation} VaR_{\alpha}(X_i,X) = - E\{ X_i| X=  
-VaR_{\alpha}(X) \} \label{VaR-CA}  
\end{equation} under some smoothness assumptions, see Tasche  
\cite{Tasche99}.  Kalkbrener et al. \cite{Kalkbrener04} report that   
capital allocation with VaR may require larger economic capital to an asset then the   
lowest possible outcome of the variable.   But then the capital  
allocation is using covariance instead of Euler allocation. This will not  
happen with Euler capital allocation as is seen from \eqref{VaR-CA}. It is much  
easier to perform capital allocation using correlation than Euler  
allocation. In Section  \ref{Sec-calc}  we show how calculation of capital  
allocation using Euler allocation may be performed efficiently with  
Markov chain Monte Carlo methods.  
  
VaR satisfies the following monotonicity property: If $P(X \leq z) \geq P(Y \leq  
z)$ for all values of $z \in R$ then $ VaR_{\alpha}(X) \geq  VaR_{\alpha }(Y) $.   
However, Euler capital allocation with VaR does not satisfy the same monotonicity   
e.g. we may have $ VaR_{\alpha}(X,X+Y) \leq  VaR_{\alpha }(Y,X+Y) $. This may be surprising.  
 Missing this property may be more   
important than missing subadditivity when considering whether the capital  
allocation inside a corporate is fair or not. To the authors knowledge, this   
property is not proved earlier. This property is due to  
the fact that VaR focuses on only one point in the  
distribution.  
  
 We  give two examples that Euler capital allocation with VaR  
 is not  monotonous. The first example has two independent discrete  
variables and the second example has stochastic variables that are  continuous and   
dependent.   \\  
  
{\bf Example 1 }  
Let $X_1$ and $X_2$ be two independent stochastic variables where  
 $$P(X_1=0)=P(X_2=0)=0.9925$$ and   
$$P(X_1=-200)=P(X_2=-100)=0.0075.$$  
Then   
\begin{align}  
  VaR_{0.99}(X_1)& = VaR_{0.99}(X_2)=0 \nonumber \\   
VaR_{0.99}(X_1+X_2) &=100 \nonumber \\  
  VaR_{0.99}(X_1,X_1+X_2) & =0 \nonumber \\  
 VaR_{0.99}(X_2,X_1+X_2) & =100 \nonumber  
  \end{align}   
Hence, the asset $X_2$ gets allocated all the risk even  
  though $ X_1\leq X_2$. This may seem surprising and the example is  
  analysed more thoroughly by introducing weights $(u_1,u_2)$ for the  
  two assets, e.g.   
$$ X(u_1,u_2)= u_1 X_1 +u_2 X_2. $$  
 Then we have  
\begin{align}   
  VaR_{0.99}(X(u_1,1)) &= 100  u_1 \mbox{ \ \ for  $0 \geq u_1 \geq 2$ }  
 \nonumber\\  
 VaR_{0.99}(X(1,u_2)) &= 100 \mbox{ \ \ for $ u_2 \geq 0.5.$ } \nonumber \end{align}  
 We see that the risk for the portfolio is sensitive to  
  weight $u_1$ but not $u_2$ in the interesting region close to $(u_1,u_2)=(1,1)$,  
 hence it makes sense to reduce the  
  weight of $X_1$ instead of $X_2$ if the regulatory requirement is  
  based on $ VaR_{0.99}(X).$ This also illustrates that it is possible  
  to cheat in the system. The person responsible for asset $X_1$ may  
  reduce the capital allocated to this asset if he promises to give a  
  value of 100 to charity if $X_1=-100.$ This makes the distribution  
  of $X_1$ and $X_2$ equal implying that they get the same capital  
  allocation.  
 \\  
  
{\bf Example 2 }  
Let $X=3X_1+X_2$ where $X_1$ is symmetric around $0$   
and  $X_2$ is given as   
$$X_2 = \left\{ \begin{array}{cc}  
-X_1 & \mbox{ if $X_1 \leq 0$ }\\  
-2 X_1 & \mbox{ if $  X_1   > 0 $. }   
\end{array} \right. $$  
 This implies that $X_2$ has the same upside as $X_1$,  
 but the downside is not as good considered as a univariate variable. But $X_2$ has good   
diversification properties in the portfolio. The portfolio $X$ has the same marginal distribution as  $X_2$,   
$$X = \left\{ \begin{array}{cc}  
2 X_1 & \mbox{ if $X_1 \leq 0$ }\\  
 X_1 & \mbox{ if $  X_1   > 0 $. }   
\end{array} \right. $$  
 Then $VaR_{\alpha} (X_2,X) < 0<   
VaR_{\alpha} (X_1,X)$  even though $P(X_1\leq z) \leq P(X_2\leq z)$ all   
$z \in R$.\\  
  
The situation is very different in the two cases. In Example 1 it is tempting to deviate from Euler capital allocation by moving capital between the involved assets in order to maintain monotonicity. In Example 2 it is important to not change the Euler capital allocation in order to honor the increased diversification.

\section{Expected shortfall, ES}  
  
ES is defined as  
 $$  ES_{\alpha}(X) = -E \{ X| X \leq -VaR_{\alpha}(X)  \}. $$  
 Acerbi et al, \cite{Acerbi02} prove that   
$$  ES_{\alpha}(X) = \frac{1}{1-\alpha} \int_{\alpha}^1    
VaR_{\tau}(X) d \tau .   $$  
ES is a coherent risk measure. Further, it is proved that  Euler allocation gives   
$$  ES_{\alpha}(X_i,X) = - E \{ X_i | X \leq - VaR_{\alpha}(X)  \} $$  
assuming sufficient smoothness. Since ES takes the average of VaR for   
$\alpha \leq \tau \leq 1$  we have  $  ES_{\alpha}(X) \leq    VaR_{\alpha}(X) $.    
It is also proved the ES is the smallest risk measure that is law invariant   
where $\rho(X) \leq  VaR_{\alpha}(X)$.   
This makes ES a good alternative to VaR.   
The property {\it law invariant } is that    
if $X,Y \in V$ and $P(X\leq z)=P( Y\leq z)$ for all $z\in R$ then   
 $\rho(X)= \rho(Y).$  
Since ES focuses on   
the entire tail and not only a quantile, it may require more realizations in a Monte Carlo estimation than VaR.

\section{Combining ES and VaR}  
  
Since ES is coherent while VaR is not, many authors recommend to use  
ES instead of VaR. We will discuss under the assumption that the  
regulator requires the use of VaR for setting the economic capital  
whether to use VaR or ES in the capital allocation.  Since VaR is not  
subadditive some authors seem to recommend to use ES for capital  
allocation. The missing monotonicity property strengthen this  
argument. However, we will argue that this is not necessarily a good  
choice. It is correct that VaR may lead to examples where we do not  
get subadditivity and therefore there may be arguments for splitting  
the portfolio in several parts. However, most practitioners report a  
30\% diversification effect indicating that the missing subadditivity  
property is mainly academic. The following example shows that  
combining VaR as a risk measure for the portfolio and ES in the  
capital allocation, may give conflict of interest. This is avoided if  
VaR or ES is used both as a risk measure for the portfolio and in  
the capital allocation. This  indicates that if VaR is used as risk  
measure, then VaR should also be used in the capital allocation.   
  
  
\begin{figure}               
\begin{center}             
\includegraphics[width=0.5\linewidth,angle=-90]{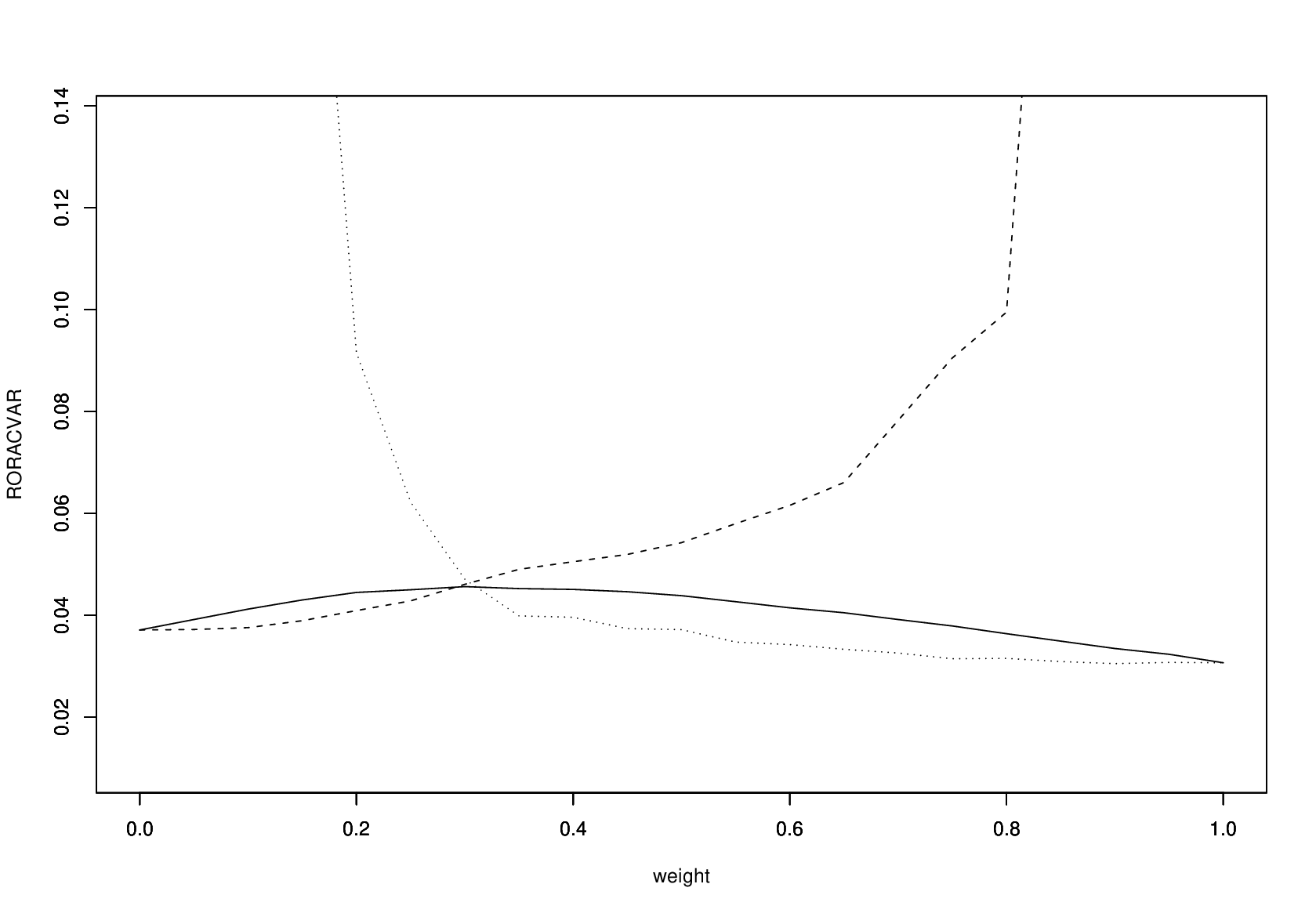}   
\caption{ RORAC for the porfolio (line) and the assets  $u X_1$ (dotted) and $(1-u) X_1$ (dashed)   
with Euler capital allocation using VaR. }   
 \label{fig_RARVaR}  
\end{center}             
 \end{figure}    
  
\begin{figure}               
\begin{center}             
\includegraphics[width=0.5\linewidth,angle=-90]{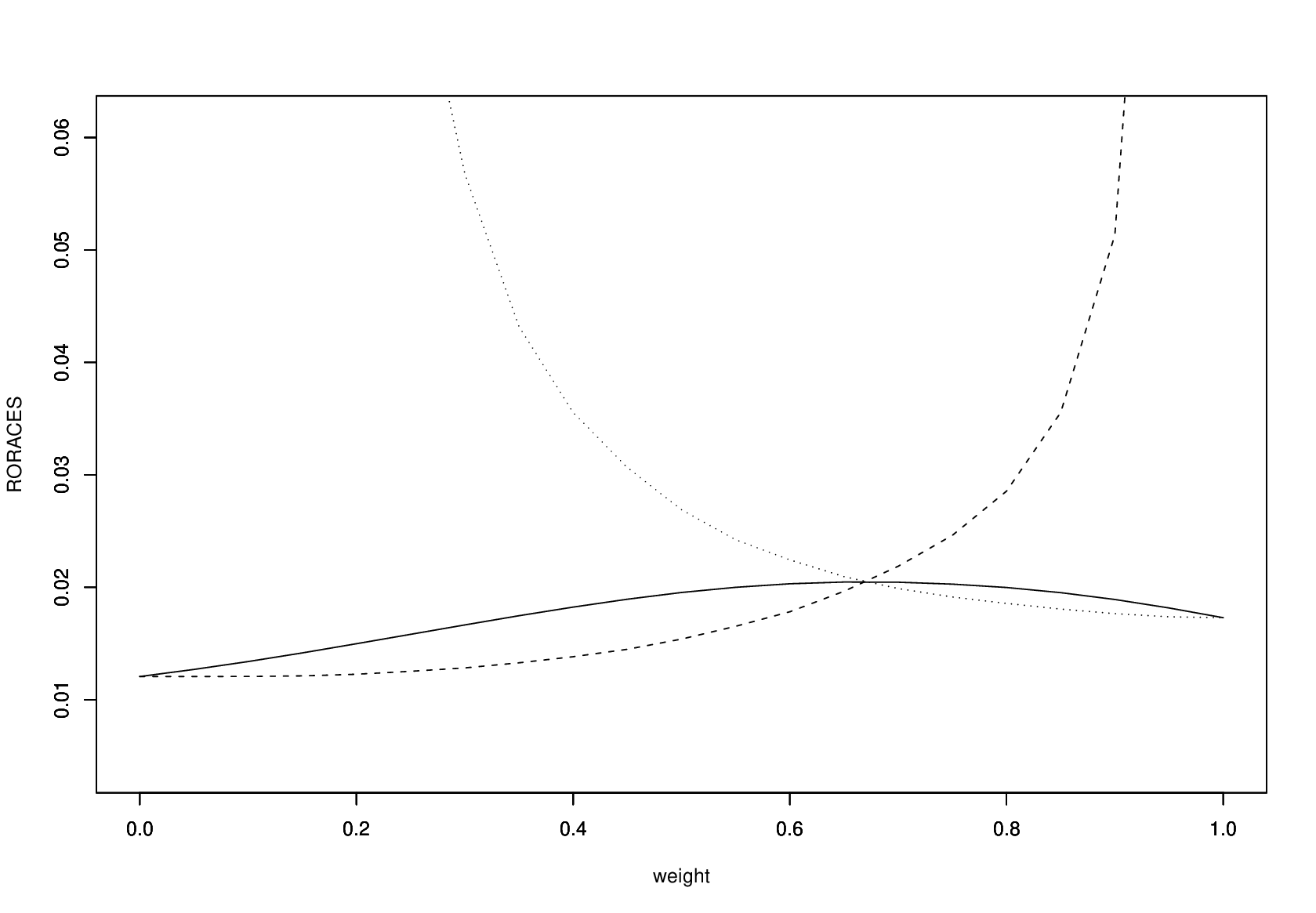}   
\caption{ RORAC for the porfolio (line) and the assets  $u X_1$ (dotted) and $(1-u) X_1$ (dashed)   
with Euler capital allocation using ES.}    
\label{fig_RARES}  
\end{center}             
 \end{figure}    
  
\begin{figure}               
\begin{center}             
\includegraphics[width=0.5\linewidth,angle=-90]{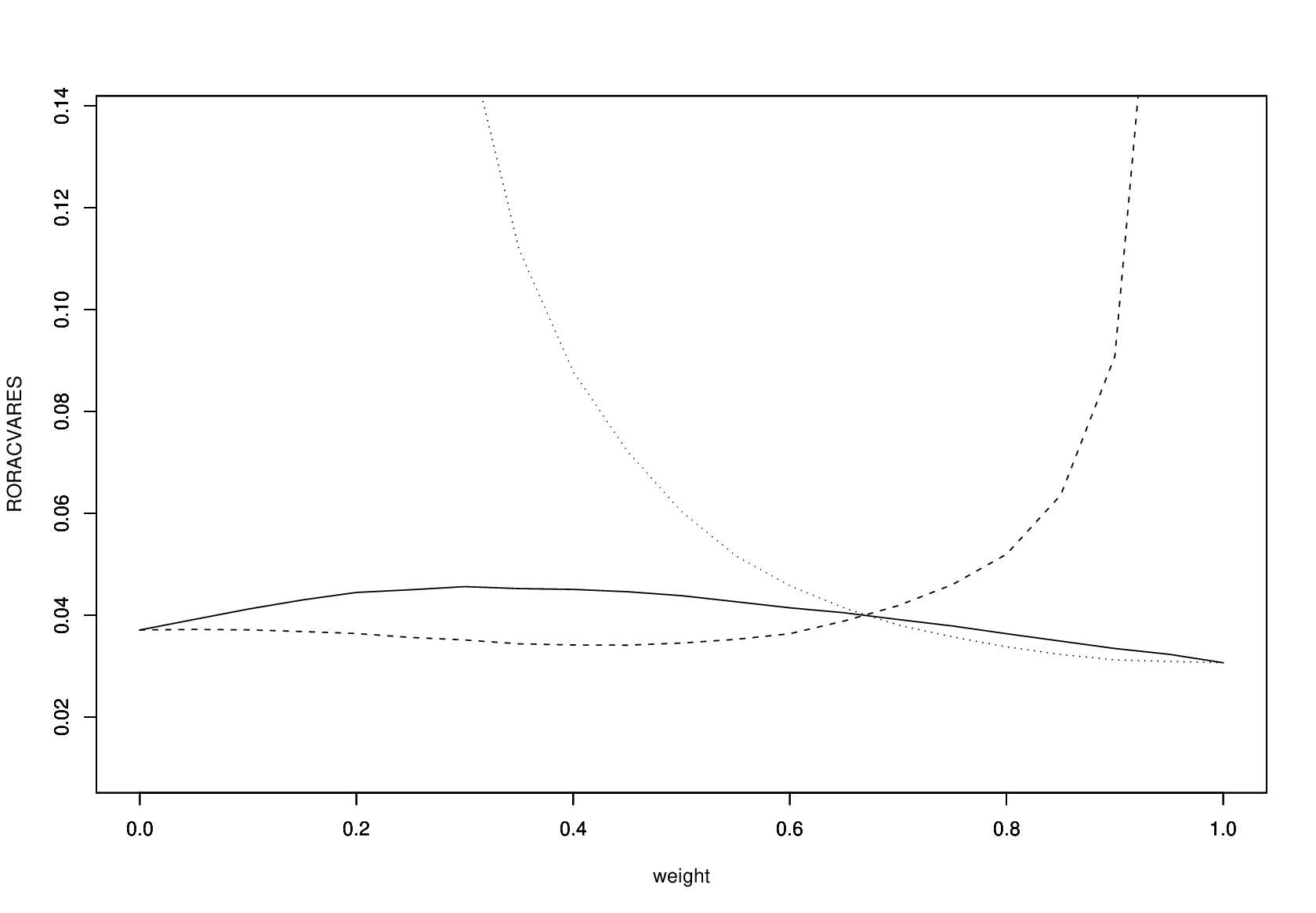}   
\caption{ RORAC for the porfolio (line) and the assets  $u X_1$ (dotted) and $(1-u) X_1$ (dashed)   
with VaR as risk measure and  $\rho_{VaR-ES}(X_i,X)$ used in the capital allocation. }  \label{fig_RARVaRES}  
\end{center}             
 \end{figure}    
  
{\bf Example 3 } Let $X_i=0.5-I_i Y_i$ for $i=1,2$ where $I_i$ is an  
indicator and $Y_i$ is a Pareto distributed variable with density   
$$  
f(y)= \frac{\gamma_i}{b_i}(\frac{y}{b_i}+1)^{-\gamma_i-1}$$  
 where  
$y>0.$ The indicator $I_i=1 $ with probability $0.1$ and else  
$I_i=0$. We have chosen the variables $\gamma_1=5$ and $\gamma_2=1.7$  
and $b_i$ such that $E\{ X_i \}=0.2$ for $i=1,2$. We study the  
portfolio $X=u X_1 + (1-u) X_2$ for $0 \leq u \leq 1$ and compare  
RORAC with Euler capital allocation using  VaR and ES  as risk  
measure (Figures  
\ref{fig_RARVaR} and \ref{fig_RARES}), and when VaR is used as risk measure for the portfolio  while  
$$ \rho_{VaR-ES}(X_i,X)= ES_{\alpha}(X_i,X)  
VaR_{\alpha}(X)/ES_{\alpha}(X)$$   
is used in the capital allocation ( Figure \ref{fig_RARVaRES}).  
The figures  show RORAC for the portfolio and  the two  
assets $u X_1$ and $(1-u) X_1$ as a function of $u$ for the different risk measures and   
 capital allocation methods.   
 When we  use Euler capital allocation   
with VaR  
or ES we get an optimal portfolio for $u=0.3$ and $u=0.7$ respectively  
where both assets and the total portfolio have the same  
RORAC. That the assets have the same RORAC in the optimum as  
the total portfolio follows from the RORAC compatibility of Euler  
capital allocation.  In these two cases RORAC is increasing  
for the assets when the asset gets less weight in the portfolio.   
 When $\rho_{VaR-ES}$ is used for the capital  
allocation the performance is more complex. The diversification effect  
is not as expected and the two assets have very different RORAC at the  
optimal diversification. If we require that the two assets have  
approximately the same RORAC then the total portfolio is far from  
optimal. If we reduce the weight of the asset with lowest RORAC, we may reduces the RORAC of the total portfolio.   
\\

 \section{Changes in the portfolio}  
  
Assume we already have a portfolio $X$ and consider to expand the  
portfolio with $X_{n+1}$.  Then we should base the evaluation on the  
capital allocation  
\begin{equation}  
\rho(X_{n+1},X+X_{n+1}). \label{new}  
\end{equation}  
 Euler capital allocation  does not encourage new investments $X_{n+1}$ where   
$$ \rho(X,X+X_{n+1})-\rho(X)$$   
is large, hence contributes to a reduction in  
the capital allocated to the rest of the portfolio. There are capital allocation  
methods that takes these other properties into account. But Tasche  
\cite{Tasche04} proves that for continuous differentiable,  
subadditive and positive homogeneous risk measures then  Euler  
allocation gives   
$$ \rho(X_{n+1},X+X_{n+1}) \geq \rho(X+X_{n+1})-\rho(X). $$   
Hence if each asset only gets allocated the additional  increase  
in the total portfolio, then the sum of the capital allocated to each  
asset may not add up to the capital required for the entire portfolio  
and the sum will never be above. In some cases it may be fair that the  
last asset, $X_{n+1}$, gets allocated only the marginal increase and  
all the other assets get allocated capital assuming that $X_{n+1}$ is  
not present. But then it is necessary to argue that this particular asset  
 is different from the rest and deserves a special treatment.   
  
In a one-period framework, as pointed out by Fischer \cite{Fischer03},  
where we at time $t=0$ can evaluate the entire portfolio $X$ and we  
are not allowed to do any changes before $t=T$ and the variables $X$  
and $X_{n+1}$ denote the values at time $t=T$, then it is natural to  
base the evaluation on \eqref{new}. Euler capital allocation is only  
based on the risk as it is evaluated today. The typical situation is  
often very different.  It does not consider that a large part of the  
portfolio may be invested a long time ago, under different economic  
conditions and that the corporate may not be in the position to remove  
part of the portfolio that is not beneficial any more.  The present  
decision is whether to include a new investment in the portfolio. Let  
$X_t$ and $X_{n+1,t}$ denote the portfolio and the new investment at  
time $t$ and assume the new investment is a commitment from $t=0$ to  
$t=T.$ Then it is natural to base the decision on the capital  
allocation   
$$ \frac{1}{T} \int_{t=0}^T E\{ \rho(X_{n+1,t},X_t+X_{n+1,t})\} dt.  $$   
In order to evaluate this  
expression it is necessary to evaluate the future properties of the  
portfolio and the new investment. This is typically done by  
simulation. The following schematic example shows how this may be  
done.

\begin{table}          
\caption{ RORAC for a new investment depending on the other investment.} \label{tab:comb}  
\begin{tabular}{|c|c|c|}  
\hline New investment & \multicolumn{2}{c|}{  Other investment } \\   
\hline               & $X_1$ & $X_2$ \\  
\hline $X_1$ &   0.030  &  0.033  \\  
\hline $X_2$ &    0.071   &  0.044 \\  
\hline \end{tabular}  
\end{table}  
\begin{table}          
\caption{ Average RORAC for a two years period for a new investment depending on the old investment in the portfolio}  
 \label{tab:port}  
\begin{tabular}{|c|c|c|}  
\hline New investment & \multicolumn{2}{c|}{  Old investment } \\   
\hline  & $X_1$ & $X_2$ \\  
\hline  $X_1$ &   0.031  &  0.032  \\  
\hline  $X_2$ &   0.064  &  0.051 \\ \hline   
\end{tabular}  
\end{table}  
  
{\bf Example 4 }   
Let the portfolio be $X=Z_1+Z_2$ where one of the  
$Z_j$ is renewed each year for a period of two years. Each of the  
variables $Z_j$ may be of the two types $X_i$, $i=1,2$ as defined in Example  
3. When we make the investment for a two years period we know the  
portfolio for the first year, but for the second year the other part  
of the portfolio is with equal probability equal to $X_1$ or $X_2$.  
Table \ref{tab:comb} shows  RORAC for the four cases.   
If the new investment is of type $X_2$ and the other investment that  
is lasting for one more year is also of type $X_2$, then the RORAC for  
the new investment the first year is 0.044. For the second year will  
the other investment be of type $X_1$ or $X_2$ with equal probability  
leading to a RORAC equal 0.071 or 0.044 with equal probability. This  
implies that the expected average annual RORAC for the new investment  
is equal to (0.044+(0.044+0.071)/2)/2=0.51. By a similar method we may  
calculate the numbers shown in Table \ref{tab:port}. This table  
 may be used in the evaluation of a new investment based   
on whether the new and the old investment is of type $X_1$ or $X_2$.

 \section{Calculation of capital allocation} \label{Sec-calc}  
  
The easiest method to calculate the Euler capital allocation based on the stochastic  
properties of the portfolio and a risk measure is to use Monte Carlo  
simulation. Then it is generated a large number of realizations and  
the risk measure is calculated by evaluating the realizations  
empirically. If VaR or ES is used as risk measure with a high value  
$\alpha$, we will only use a small part of the simulated values. This  
makes the simulation inefficient. This may be considerably improved by  
using importance sampling, see e.g. Kalkbrener et al. \cite{Kalkbrener04} or Tasche \cite{Tasche06}.  
 If we use  
Markov chain Monte Carlo (MCMC) it is possible to improve the sampling even  
more. See Meyn and Tweedie \cite{Meyn93} for a thorough introduction to MCMC.   
In the following  example we illustrate how capital allocation may be  
calculated using Monte Carlo simulation, Importance sampling and  
MCMC. We give complete algorithms for each  
type. It is outside the scope of this paper to discuss the  
different alternatives within each class in detail.

{\bf Example 5 }  
 Let $$X^j=  
\sum_{i=1}^n X^j_{i}$$ denote realization number $j$ of the  
portfolio. Assume $$ X^j_{i} = a_i - exp(Y^j_{i}) $$ where $ Y^j_{i}  
\sim N(\mu_{i},\sigma_i^2) $  and all $ Y^j_{i}$ are uncorrelated in order to simplify the notation.  
  Let $X^{(j)}$ denote the $X^{j}$ sorted such that $$X^{(1)} \leq  
X^{(2)} \leq \cdots \leq X^{(n)} $$ and $X^{(j)}_{i}$ is the realization  
for asset $i$ that correspond to $X^{(j)}$.  Assume $n_{\alpha  
}=(1-\alpha) n$ is an integer such that $-X^{( n_{\alpha })}$ is a  
reasonable estimator for $ VaR_{\alpha}(X)$. Let $b$ be an integer  
such that $ b \leq n_{\alpha } $.

The Monte Carlo simulation algorithm below gives n realizations of $X$. From these realizations we find
 the estimators   
$ VaR_{\alpha, MC }(X) $ for $ VaR_{\alpha}(X)$ and $ VaR_{\alpha, MC }(X_i,X) $ for $ VaR_{\alpha}(X_i,X)$.  
\begin{enumerate}  
\item For $j=1,2, \cdots , m$  
\begin{enumerate}  
\item  Set $X^j= \sum_{i=1}^n (a_i - exp(Y^j_{i})) $  
\end{enumerate}  
\item Sort to find $X^{(1)},\cdots , X^{(m)},$ and the corresponding $X^{(j)}_i$  
\item Set $ VaR_{\alpha, MC }(X) = -X^{( n_{\alpha })} $   
\item Set $ VaR_{\alpha, MC }(X_i,X) = - \frac{1}{2b+1}   
X^{( n_{\alpha })} \sum_{j=-b}^b \frac{X^{ (n_{\alpha }+j)}_{i}}  
{X^{ (n_{\alpha }+j})} $  
\end{enumerate}  
  
In order to make good estimates for the capital allocation, it is  
necessary with many realizations estimating this, hence $b$ large. On  
the other hand it is necessary with $n$ large such that the $2b+1$  
realizations $ X^{( n_{\alpha }-b)}, \cdots , X^{( n_{\alpha }+b)}$  
 all are sufficiently close to the quantile $ VaR_{\alpha}(X)$.   
The storage problem may be solved by only storing  
realizations in an interval $ d_1 \leq X^{j} \leq d_2 $ surrounding $  
VaR_{\alpha}(X)$.  
  
Only a small portion of the realizations is in the interesting  
interval. This may be improved by importance sampling. Here is a very simply  
importance sampling algorithm where $Y^j_{i,IS} \sim N(\mu_{i,IS},\sigma_{i,IS}^2)  $.  
   Define  $\phi_{\mu,\sigma^2}(y)$ as  the density in   
the $N(\mu,\sigma^2)$ distribution and $p^j$ the importance sampling weight for each realization.   
Further, define $p^{(j)}$ as  
the values of $p^j$ sorted according to the size of $X^j $ and scaled  
such that $0=p^{(0)}< p^{(1)} < \cdots < p^{(m)}=1$ and $p^{(j+1)}  
-p^{(j)}=c p^k$ where $p^k$ is the corresponding value before the  
sorting. Define  $n_{\alpha,IS }$ such that  
$p^{(n_{\alpha,IS })}$ is closest to $(1-\alpha)n$. We may then define the importance sampler estimators  
$ VaR_{\alpha, IS}(X) $ and $ VaR_{\alpha, IS }(X_i,X) $ by the algorithm:  
\begin{enumerate}  
\item For $j=1,2, \cdots , m$  
\begin{enumerate}  
\item  Set $X^j= \sum_{i=1}^n (a_i - exp(Y^j_{i} )) $  
\item  Set $p^j= \prod_{i=1}^n \frac{\phi_{\mu_{i},\sigma_i^2}(Y^i_{j})}{  
                                     \phi_{\mu_{i,IS},\sigma_{i,IS}^2}(Y^i_{j})} $  
\end{enumerate}  
\item Sort to find $X^{(1)},\cdots , X^{(m)},$ and the corresponding $X^{(j)}_i$ and $p^{(j)}$.  
\item Set $ VaR_{\alpha, MC }(X) = -X^{( n_{\alpha, IS })} $.   
\item Set $ VaR_{\alpha, MC }(X_i,X) = - \frac{1}{2b+1}   
X^{( n_{\alpha,IS })} \sum_{j=-b_{IS}}^{b_{IS}} \frac{X^{ (n_{\alpha,IS }+j)}_{i}}  
{X^{ (n_{\alpha }+j})} $.  
\end{enumerate}  
  
Still, only a small part of the realizations are used in the  
estimation and none of these are exactly such that $X= VaR_{\alpha  
}(X). $ By using Markov chain Monte Carlo MCMC it is possible to make  
all realizations equal to $X= VaR_{\alpha }(X) $ and such that all  
realizations may be used in the capital allocation. Assume  
$VaR_{\alpha }(X) $ is found by another algorithm, for example one of  
the two algorithms described above. It is necessary with significant  
larger $n$ in order to find the capital allocation than VaR hence it  
make sense to use on of the other algorithms for this.  Then the MCMC  
estimator for the capital allocation $ VaR_{\alpha, MCMC }(X_i,X)$ may  
be found by the following algorithm:  
  
\begin{enumerate}  
\item Let $X^0$ be  any realization satisfying $X^0= VaR_{\alpha }(X) $. 
\item Let $k_1\leq n$ be any index   
\item For $j=1,2, \cdots , m*n$  
\begin{enumerate}  
\item  Find index $k_2 \leq n$ and $k_1 \neq k_2$ uniformly  
\item  Find $\tilde{Y}^{j+1}_{ k_2 } \sim N(\mu_{k_2 },\sigma_{k_2}^2) $ and   
$corr(\tilde{Y}^{j+1}_{ k_2 },Y^{j}_{ k_2 }) $ constant  
\item  Set $\tilde{X}^j_{ k_2 }= (a_{ k_2 } - exp(Y^j_{k_2 })) $  
\item Set $\tilde{X}^j_{ k_1 }=X^j_{ k_1 }+X^j_{ k_2 }-\tilde{X}^j_{ k_2 }$  
\item  Set $\tilde{Y}^j_{ k_1 }= \log(a_{ k_1 }- \tilde{X}^j_{k_1 }) $  
\item  Set $p_j=  \phi_{\mu_{i},\sigma_i^2}(\tilde{Y}^i_{j})/\phi_{\mu_{i},\sigma_i^2}(Y^i_{j}) $  
\item Set $X^{j+1}=\tilde{X}^{j+1}$ with probability $\min \{1,p_j    \} $ and else set $X^{j+1}=X^{j}$.  
\item Set $k_1=k_2$  
\end{enumerate}  
\item   
Set $ VaR_{\alpha, MCMC }(X_i,X) = - \frac{1}{m}  \sum_{j=1}^m X^{ j n}_{i} $.  
\end{enumerate}  
  
The MCMC algorithm generates a chain of realizations  $X^j$ that are according to the   
distribution we want to study.  
 But the first realizations in the chain are not from the distribution (a burn-in period)  and   
 realizations  $X^j$ and  $X^{j+k}$ are dependent but the dependence decreases geometrically in $k$.  
Table  \ref{tab: operat} shows the number of numerical operations  for the different algorithms.  
 Notice that the number of operations is considerably  
smaller for MCMC in the capital allocation per realization.    
  
\begin{table}          
\caption{ Number of numerical operations (additions, multiplications, exponential, logarithm ) per realization   
in the three algorithms }  
  \label{tab: operat}       
\begin{tabular}{|c|c|c|c|}  
\hline & MC  & IS & MCMC \\  
\hline $VaR(X)$ &    3n &   6n  &   -   \\  
\hline $VaR(X_i,X)$ &  $3n/(2b+1)$ &  $6n/(2b_{IS}+1)$ &  9  \\  
\hline \end{tabular}  
\end{table}

We have tested the three algorithm in an example with $n=90$ assets with the following parameters:   
 $\sigma_i=0.5 $ and $  E \{ X_{i} \} =0.2$ for all $i$  while the assets are divided into three groups with  
30 assets in each group where $\mu_{i}=0.44/0.45/0.47$ respectively.   
The parameters $a_i$ are determined from the other parameters.   
Table  \ref{tab:estim}  shows the  estimate for  $VaR_{0.99}(X)$ and the capital allocation   
$VaR_{0.99}(X_i,X)$ for the portfolio for the three different algorithms.   
 By having 30 equal assets in the portfolio, it is possible to estimate the precision of the different estimates.   
\begin{table}          
\caption{ Estimate for the VaR and the capital allocation using the three algorithms. The standard deviation of the  
estimates are given en parenthesis.  }  
  \label{tab:estim}        
\begin{tabular}{|c|c|c|c|}  
\hline & MC  & IS & MCMC \\  
\hline m                           &    1.000.000 &   1.000.000  &   100.000   \\  
\hline b                           &  1.600   &  20.000 &  -  \\  
\hline  $VaR_{0.99}(X)$            &  6.33  &  6.38 &  -  \\  
\hline  $VaR_{0.99}(X_1,X)$        &  0.038 (0.018) &  0.042 (0.016) &  0.038 (0.027)   \\  
\hline  $VaR_{0.99}(X_{31},X)$     &  0.064 (0.021) &  0.065 (0.016) &  0.066 (0.029)   \\  
\hline  $VaR_{0.99}(X_{61},X)$     &  0.109 (0.019) &  0.108 (0.021) &  0.109 (0.026)  \\  
\hline \end{tabular}  
\end{table}

In importance sampling we have used $\mu_{i,IS}=\mu_{i}+0.2$  
resulting in about 10 times as many realizations in an interval close  
to $VaR_{0.99}(X)$ as in Monte Carlo simulation. This implies that  
importance sampling may give an improvement with a factor 10 compared  
to Monte Carlo simulation. In MCMC we have used correlation equal to 0.3 when proposing a new  
$\tilde{Y}^{j+1}_{ k_2 } $ value leading to  
an acceptance ratio at 0.57 and that realizations with $X^j$ and  
$X^{j+5}$ are almost independent. There is almost no burn-in in this MCMC algorithm.   
Importance sampling gives 2b=40.000 samples from m= 1 mill. MCMC gives 100.000 samples using about 10 \% of  
the number of operations. This indicates that that MCMC may give an  
improvement with a factor 100 compared to Monte Carlo simulation.  
In the test, we have not performed any optimization of the algorithm and  parameters. Hence, it is reasonable to   
assume that it is possible to improve both importance sampling and MCMC even further.

 \section{Concluding remarks}  
  
In this paper we have given an overview over capital allocation using the Euler formula. We have contributed to the   
research showing that:  
\begin{itemize}  
\item   
Euler capital allocation with VaR is not monotone even though VaR is monotone. This is shown in both  
an example with independent  discrete variables and with continuous correlated variables.   
\item The same risk measure should be used for the portfolio and  in the capital allocation.    
\item Simulation may be used in estimating the expected RORAC over the commitment period of  an asset.   
\item Markov chain Monte Carlo may be used in the estimation of the capital allocation.   
 \end{itemize}  
  

\end{document}